# Effective Field Theory for Distorted Photonic Crystals: Exact Solutions of the Geodesic Equation

**Hitoshi Kitagawa[1, *], Kanji Nanjyo[1], Kyoko Kitamura[1, 2,*]**

[1.] Department of Electronics, Kyoto Institute of Technology, Matsugasaki, Sakyo-ku, Kyoto 606-8585, Japan

[2.] Japan Science and Technology Agency, Precursory Research for Embryonic Science and Technology, 4-1-8 Honcho, Kawaguchi, Saitama, 332-0012, Japan

E-mail:

*pamkitag@kit.ac.jp

## ABSTRACT

We used differential geometry to develop an effective field theory to study the behavior of light propagation in distorted photonic crystals (D-PCs). To study D-PC light-ray trajectories, we derived a geodesic equation based on the principle of least action by defining the metric tensor in terms of the lattice-position distortion. The geodesic equation implies that lattice-position distortion can curve the trajectory. We present some exact solutions for the trajectory in the case of simple distortion and show that these results agree well with the simulated finite-difference time-domain results.

# I. INTRODUCTION

Einstein's geometrization philosophy for general relativity [1–4] is a classic example of the geometrization of physical laws; it has also led to the development of theoretical physics. There are close relationships between topological ideas and modern quantum field theory [5–9] or topological insulators [10,11]. Transformation optics [12,13] are also constructed based on differential geometry.

Photonic crystals (PCs) have been studied as the arrangement of well-ordered, periodic lattice points. The periodic structure facilitates the peculiar photonic dispersion that controls light behavior in various media. The modification of lattice-point positions within a range that is more narrow than the periodicity has also been applied to enhance existing nano-cavity effects [14, 15], or to add diffraction effects to the resonant mode [16, 17]. There are also some precedent studies on PCs that consisted of lattices without a periodic structure (or breaking periodicity); these studies entailed applying Newtonian mechanics to wave packets of light to enhance the optical Hall effect [18, 19]. Similarly, Deng et al. [20] reported on a distorted photonic graphene structure in which the light is bent by the effective magnetic field produced by the Dirac point in the photonic band. However, to the best of our knowledge, there are few studies that involve applying differential geometry to PC structures.

The key concepts of this theory are as follows. The eigenstates of light in PCs are Bloch states. If random or steep fluctuations are applied, the Bloch states should be scattered. However, if the rate of fluctuations is sufficiently low, the Bloch states may be changed adiabatically. Based on this physical concept, we postulated that the regular structure reflects a flat space-time, whereas the gradual changes of the structure reflect a curved space-time. Table I outlines the similarities between this theory and general relativity. The related discussion on gravity arising from lattice distortion is presented in a subsequent section.

In this manuscript, we discuss the characteristics of distorted photonic crystals (D-PCs), i.e., the gradual spatial distortion that occurs in PC structures, from the perspective of differential geometry in order to elucidate the behavior of light passing through D-PCs. Thus, we first construct an effective field theory for D-PCs. Then, we show that this theory can be utilized to develop a geodesic equation to describe the trajectory of the light ray. Finally, the exact solution is applied to simple examples, and the results are compared to those of a finite-difference time-domain (FDTD) simulation.

## II.    Effective Field Theory for Distorted Photonic Crystals

First, let us begin by defining the lattice-distortion tensors. In this theory, we assume the lattice vectors; therefore, the base vectors of the lattice correspond to the gradual function of space points. Considering the lattice-point displacements $\Delta \mathbf{r}(\mathbf{x})$ in the two-dimensional (2-D) square lattice (Fig. 1), the diagonal elements of the lattice-distortion tensors can be defined as

$$\Delta r_1(x_1 + a^{(0)}, x_2) - \Delta r_1(x_1, x_2) \equiv \Delta^L_{11}$$
$$\Delta r_2(x_1, x_2 + a^{(0)}) - \Delta r_2(x_1, x_2) \equiv \Delta^L_{22}, \quad (1)$$

whereas the off-diagonal elements of lattice-distortion tensors can be defined as

$$\Delta r_2(x_1 + a^{(0)}, x_2) - \Delta r_2(x_1, x_2) \equiv \Delta^L_{12}$$
$$\Delta r_1(x_1, x_2 + a^{(0)}) - \Delta r_1(x_1, x_2) \equiv \Delta^L_{21}, \quad (2)$$

where $a^{(0)}$ is the undistorted lattice constant.

Next, to consider the dielectric properties of the PC, we define the distortion tensor, i.e., the renormalized lattice-distortion tensor $\Delta_{ij}$, as

$$\Delta_{ij} \equiv \gamma \cdot \Delta^L_{ij}. \quad (3)$$

Here, $\gamma$ is the normalized weight factor, which is defined as follows:

$$\gamma \equiv [\frac{\sum_{\mathbf{G} \neq 0} |\kappa(\mathbf{G})|^2}{(\sum_{\mathbf{G} \neq 0} |\kappa(\mathbf{G})|^2)^{Max}}]^{(0)} \frac{\sigma_P + 1}{2}, \quad (4)$$

$$\frac{1}{\varepsilon(\mathbf{r})} = \sum_{\mathbf{G}} \kappa(\mathbf{G}) \exp(i\mathbf{G} \cdot \mathbf{r}), \quad (5)$$

where $\sigma_p$ = 1 for *H*-polarization, $\sigma_p$ = -1 for *E*-polarization, and $\varepsilon(\mathbf{r})$ is the periodic dielectric function of undistorted (or regular) PCs. (Note that in this study, the dielectric constant of media was set to be $3.5^2$.) As explained via the equations, $\gamma$ contains the following information about undistorted PCs: the filling factor, the dielectric contrast between the lattice point and background, and the inclusion of the second-order perturbative corrections of the eigenvalues for PCs. Equation 4 also states that $\gamma = 0$ under the conditions of $E$-polarization. This is because the equi-frequency contour for $E$-polarization in a rectangular lattice is circular and not elliptical at long-wavelength frequencies and because the $E$-polarized light is not affected by lattice distortion.

### III. Light propagation in D-PCs

#### A. METRIC TENSOR

We defined the D-PC field in the previous section. Here, we will discuss illuminating this field. Our ansatz is that D-PCs distorts space-time in a way that can be observed by analyzing light transmission. In consideration of this ansatz, we will consider the metric tensor as a means to evaluate the space-time structure of D-PCs. The relationship between $\mathbf{e}_i^{(0)}$ and $\mathbf{e}_j$ ($i, j = 1,2$) is represented in terms of distortion tensors as follows:

$$\begin{bmatrix} \mathbf{e}_1 \\ \mathbf{e}_2 \end{bmatrix} = \begin{bmatrix} 1 + \Delta_{11}/a^{(0)} & \Delta_{12}/a^{(0)} \\ \Delta_{21}/a^{(0)} & 1 + \Delta_{22}/a^{(0)} \end{bmatrix} \begin{bmatrix} \mathbf{e}_1^{(0)} \\ \mathbf{e}_2^{(0)} \end{bmatrix}, \qquad (6)$$

where $\mathbf{e}_i^{(0)}$ and $\mathbf{e}_i$ ($i = 1,2$) are the undistorted and distorted base vectors, respectively. The metric tensor of space is constructed by incorporating the sets of scalar products of the base vectors. Specifically, upon considering time and the refractive indices, the metric tensor of space-time can be constructed as follows:

$$g_{\mu\nu} = \begin{bmatrix} -1 & 0 & 0 \\ 0 & (n^{(0)})^2 \{1 + 2(\Delta_{11}/a^{(0)} + \Delta n(\mathbf{x})/n^{(0)})\} & (n^{(0)})^2 (\Delta_{12} + \Delta_{21})/a^{(0)} \\ 0 & (n^{(0)})^2 (\Delta_{12} + \Delta_{21})/a^{(0)} & (n^{(0)})^2 \{1 + 2(\Delta_{22}/a^{(0)} + \Delta n(\mathbf{x})/n^{(0)})\} \end{bmatrix}$$

(7)

where $n^{(0)}$ and $\Delta n(\mathbf{x})$ are the average and deviated refractive indices of the D-PC,

respectively. This metric tensor possesses all the geometric information about the D-PC.

## B. GEODESIC EQUATION

Next, we consider the trajectory of light by developing the geodesic equation. The action of a D-PC is given by

$$S = \int \sqrt{-g_{\mu\nu} \frac{dx^\mu}{d\sigma} \frac{dx^\nu}{d\sigma}} d\sigma, \qquad (8)$$

where $\sigma$ is a Lorentz invariant parameter, and $\mu, \nu = 0,1,2$. Then, the principle of least action, i.e.,

$$\delta S = 0 \qquad (9)$$

can be utilized to develop the geodesic equation, which is given by

$$\frac{d^2 x^\mu}{d\tau^2} + \Gamma^\mu_{\nu\lambda} \frac{dx^\nu}{d\tau} \frac{dx^\lambda}{d\tau} = 0, \qquad (10)$$

where $\Gamma^\mu_{\nu\lambda}$ is the connection coefficient given by the metric tensor of the D-PC and is given by

$$\Gamma^\mu_{\nu\lambda} = g^{\mu\tau} \frac{1}{2} (\partial_\lambda g_{\tau\nu} + \partial_\nu g_{\tau\lambda} - \partial_\tau g_{\nu\lambda}). \qquad (11)$$

The elements of the connection coefficient are given by

$$\Gamma^0_{\nu\lambda} = 0,$$

$$\Gamma^1_{11} = \partial_1 \left(\frac{\Delta_{11}}{a^{(0)}} + \frac{\Delta n}{n^{(0)}}\right), \Gamma^1_{12} = \partial_2 \left(\frac{\Delta_{11}}{a^{(0)}} + \frac{\Delta n}{n^{(0)}}\right), \Gamma^1_{22} = \partial_2 \left(\frac{\Delta_{12} + \Delta_{21}}{a^{(0)}}\right) - \partial_1 \left(\frac{\Delta_{22}}{a^{(0)}} + \frac{\Delta n}{n^{(0)}}\right), \qquad (12)$$

$$\Gamma^2_{22} = \partial_2 \left(\frac{\Delta_{22}}{a^{(0)}} + \frac{\Delta n}{n^{(0)}}\right), \Gamma^2_{21} = \partial_1 \left(\frac{\Delta_{22}}{a^{(0)}} + \frac{\Delta n}{n^{(0)}}\right), \Gamma^2_{11} = \partial_1 \left(\frac{\Delta_{12} + \Delta_{21}}{a^{(0)}}\right) - \partial_2 \left(\frac{\Delta_{11}}{a^{(0)}} + \frac{\Delta n}{n^{(0)}}\right).$$

## C. EXACT SOLUTIONS FOR SIMPLE MODELS

We consider the simple cases in which the local displacement $\Delta\mathbf{r}(\mathbf{x})$ is given as the quadratic function of the space points. In these cases, the geodesic equation provides exact solutions. Illustrations of the models and solutions are shown in Fig. 2(a)–(b). Fig. 2(a) shows uniaxial distortion, and Fig. 2(b) shows biaxial distortion. To only consider the effects of lattice distortion, i.e., $\Delta n(\mathbf{x}) = 0$, we changed the radii $r^{(m, n)}$ in the direction of the distortion to ensure that all of the filling factors for the lattice point in the unit cell had the same values. In the case of uniaxial distortion, the displacement $\Delta \mathbf{r}(\mathbf{x})$ can be given by

$$\Delta \mathbf{r}(\mathbf{x}) / a^{(0)} = (0, \beta(y/a^{(0)})^2) \qquad (13)$$

where the constant $\beta$ is a dimensionless distortion coefficient for the $y$ direction. We set $r^{(n)} = r^{(0)} \sqrt{1 + 2n\beta}$, where $r^{(0)}$ is the lattice-point radius of the square lattice PC. In this case, the geodesic equation is given by

$$\frac{d^2 x}{d\sigma^2} = 0,$$
$$\frac{d^2 y}{d\sigma^2} = -2\gamma\beta(\frac{dy}{d\sigma})^2. \qquad (14)$$

The solution of these simultaneous differential equations, which yields the trajectory of the light ray input from the origin, is given by

$$y = \frac{1}{2\gamma\beta} \ln|(2\gamma\beta \tan\phi)x + 1| \qquad (15)$$

where the incident angle from the $x$ axis is $\phi$. Figure 2(c) shows the trajectories of the light ray in the cases of $\beta = \pm 0.006$. Note that the sign of $\beta$ changes the bending direction.

In the case of biaxial distortion, the displacement $\Delta \mathbf{r}(\mathbf{x})$ is given by

$$\Delta \mathbf{r}(\mathbf{x}) / a^{(0)} = (\alpha(x/a^{(0)})^2, \beta(y/a^{(0)})^2) \qquad (16)$$

where $\alpha$ and $\beta$ are the dimensionless distortion coefficients for the $x$ and $y$ directions, respectively. After setting $r^{(m,n)} = r^{(0)} \sqrt{(1 + 2m\alpha)(1 + 2n\beta)}$, the geodesic equation can be expressed as follows:

$$\frac{d^2x}{d\sigma^2} = -2\gamma\alpha(\frac{dx}{d\sigma})^2,$$
$$\frac{d^2y}{d\sigma^2} = -2\gamma\beta(\frac{dy}{d\sigma})^2. \qquad (17)$$

Then, the trajectory of the light ray is defined as follows:

$$y = \frac{1}{2\gamma\beta}\ln\left|(\frac{\beta}{\alpha}\tan\phi)(e^{2\gamma\alpha x}-1)+1\right|. \qquad (18)$$

The predicted light-ray trajectories in these cases are shown in Fig. 2(d). The hatched area indicates that the lattice points are connected to each other under the condition $\Delta n(\mathbf{x}) = 0$. In the case of $\alpha = +0.006$ and $\beta = +0.006$, the trajectory is a straight line; in contrast, the conditions of $\alpha = -0.006$ and $\beta = +0.006$ yield a strongly curved trajectory. This curved trajectory effect is very similar to that corresponding to uniaxial distortion with twice the value of $\beta$.

## IV. VERIFICATION

To verify our above-mentioned ansatz, we compared the aforementioned results to those derived via FDTD simulation. *H*-polarization was applied in the FDTD simulation, and the normalized frequency was set at 0.1. It is worth mentioning that, in this study, we focused on the behavior of light that is not under the influence of the Dirac point, in contrast to Ref. [20]; this means that the long-wavelength approximation (i.e., that the normalized frequency is less than approximately 0.1) is applicable.

Figure 3 shows a comparison of the results. The red line represents the theoretical results, and the wavy field distribution reflects the z-component of the magnetic field in the simulation. We applied $\beta = +0.006$, $\phi = \pi/4$, and $r^{(0)} = 0.4a^{(0)}$. Although the beam divergence slightly differed, in the cases of the air-hole (Fig. 3(a)) and dielectric-rod (Fig. 3(b)) lattices, the theoretical results agreed well with the simulated results. Increasing $\beta$ to +0.01 resulted in a strongly curved trajectory (Fig. 3(c)). Conversely, reducing $\beta$ significantly reduced the curvature of the trajectory, as shown in Fig. 3(d). Moreover, reducing $r^{(0)}$ to $0.2a^{(0)}$ also significantly reduced the curvature, with $r^{(0)} = 0$ tending toward a straight line (Fig. 3(e)). These findings confirm that $\gamma$ contains information about the PC. Even in the biaxial distortion case (Fig. 3(f)), the theoretical results were in agreement with the simulated results. Thus, the ansatz presented in Section III is valid.

## V.  DISCUSSION

Thus far, we have only applied $\Delta n(\mathbf{x}) = 0$ to focus solely on the effects of lattice distortion. Thus, in this section, we will explore the influence of $\Delta n(\mathbf{x})$ by setting the radii of all lattice points to be constant, i.e., $r^{(n)} = r^{(0)}$. In the uniaxial distortion case, the geodesic equations become

$$\frac{d^2 x}{d\sigma^2} = -2\partial_y \left(\frac{\Delta n}{n^{(0)}}\right)\left(\frac{dx}{d\sigma}\right)\left(\frac{dy}{d\sigma}\right),$$
$$\frac{d^2 y}{d\sigma^2} = -2\gamma\beta\left(\frac{dy}{d\sigma}\right)^2 + \left[\left(\frac{dx}{d\sigma}\right)^2 - \left(\frac{dy}{d\sigma}\right)^2\right]\partial_y \left(\frac{\Delta n}{n^{(0)}}\right), \quad (19)$$

where $\beta > 0$. Here, we set

$$\partial_y(\frac{\Delta n}{n^{(0)}}) \equiv \gamma\alpha', \tag{20}$$

and assume that

$$\alpha' \cong \alpha \tag{21}$$

and

$$\phi \cong \pi/4. \tag{22}$$

Then, incorporating Eq. (22) results in the following:

$$\begin{aligned}(\frac{dx}{d\sigma}) &\cong (\frac{dy}{d\sigma}) \\ (\frac{dx}{d\sigma})(\frac{dy}{d\sigma}) &\cong (\frac{dx}{d\sigma})^2 \\ (\frac{dx}{d\sigma})^2 - (\frac{dy}{d\sigma})^2 &\cong 0,\end{aligned} \tag{23}$$

which yields

$$\begin{aligned}\frac{d^2x}{d\sigma^2} &\cong -2\gamma\alpha(\frac{dx}{d\sigma})^2, \\ \frac{d^2y}{d\sigma^2} &\cong -2\gamma\beta(\frac{dy}{d\sigma})^2.\end{aligned} \tag{24}$$

These equations are essentially the same as those corresponding to the biaxial distortion case of $\Delta n(\mathbf{x}) = 0$. The trajectory is also given by

$$y \cong \frac{1}{2\gamma\beta}\ln\left|(\frac{\beta}{\alpha}\tan\phi)(e^{2\gamma\alpha x}-1)+1\right|. \tag{25}$$

Based on Eqs. (20) and (21), if the lattice points are air holes, then $\alpha < 0$, but if the lattice points correspond to a dielectric rod, then $\alpha > 0$. Thus, for an air-hole lattice, $\alpha > 0$ and $\beta > 0$; these conditions essentially yield the results presented as the black line in Fig. 2(d), which shows minimal light trajectory curvature. Alternatively, under the conditions of a dielectric-rod lattice, $\alpha < 0$ and $\beta > 0$; these conditions essentially yield the results presented as the red line in Fig. 2(d), which shows a strongly curved light-ray trajectory.

## VI.　SUMMARY

From the perspective of differential geometry, we have constructed an effective field theory to theoretically describe D-PCs, which have been defined as PC structures that are subject to gradual spatial distortion. According to this theory, the trajectory of a light ray can be described by a geodesic equation that is similar to that for general relativity. Our results, which were derived under the condition that the average refractive index of a D-PC is constant, indicate that the trajectory of a light ray can only be curved by introducing lattice distortion. This finding was somewhat surprising considering Fermat's principle. However, the validity of this theory was confirmed by performing FDTD simulations. In summary, we have constructed the phenomenological low-energy effective field theory for D-PCs.

## VII.　ACKNOWLEGEMENT

This work was supported by JST, PRESTO Grant Number JP 20345471, Japan. We are also grateful to Prof. S. Iwamoto, Dr. Y. Ota, Dr. Y. Nakata, Dr. M. Fujita, and Prof.

S. Noda for their valuable contributions to the discussions.

## Table & Figure Captions

**Table I**

Similarities between this theory and general relativity, where $g_{\mu\nu}$ indicates metric tensor and $ds^2 = g_{\mu\nu}dx^{\mu}dx^{\nu}$.

**Fig. 1**

Lattice distortion of a 2-D square lattice. $\boldsymbol{a}_i^{(0)}$ ($i = 1,2$) corresponds to the undistorted lattice vectors, and $\boldsymbol{a}_i$ ($i = 1,2$) corresponds to the distorted lattice vectors. $\Delta \mathbf{r}(\mathbf{x})$ is the lattice-point displacement. The lattice-distortion tensors $\Delta^L_{mn}$ ($m, n = 1, 2$) were derived based on the deviation between $\boldsymbol{a}_i^{(0)}$ and $\boldsymbol{a}_i$.

**Fig. 2**

Illustrations of example models and exact solutions. (a) Uniaxial distortion model. (b) Biaxial distortion model. (c) Light-ray trajectories in the case of uniaxial distortion. The brown, blue, and green curves respectively correspond to $\beta = +0.006$, $-0.006$, and $+0.012$. (d) Light-ray trajectories in the case of biaxial distortion. In the case of $\alpha = +0.006$ and $\beta = +0.006$, the trajectory is a straight line (black), whereas the conditions of $\alpha = -0.006$ and $\beta = +0.006$ yielded a strongly curved trajectory (red). The extent of curvature was very similar to that predicted for uniaxial distortion with a two-times larger $\beta$ value (green). The hatched area indicates that the lattice points were connected to each other under the condition of $\Delta n(\mathbf{x}) = 0$. The incident angle, $\phi$, from the x axis was $\pi/4$; the radius of the circular lattice point was set as $r^{(0)} = 0.4a^{(0)}$; $\Delta n(\mathbf{x}) = 0$.

**Fig. 3**

Comparison of the theoretical and simulated results. The red curves represent the theoretical results, and the wavy field distributions show the results of FDTD simulation. $\Delta n(\mathbf{x}) = 0$ and $\phi = \pi/4$. (a)–(e): Uniaxial distortion results. The radius of the circular lattice point was set as $r^{(0)} = 0.4a^{(0)}$ for (a)–(c). (a) Air-hole lattice results for $\beta = +0.006$. (b) Dielectric-rod lattice results for $\beta = +0.006$. (c) Air-hole lattice results for $\beta = +0.01$. (d) Air-hole lattice results for $\beta = +0.002$. (e) Air-hole lattice results for $r^{(0)} = 0.2a^{(0)}$ and $\beta = +0.006$. (f) Biaxial distortion results for $\alpha = -0.006$, $\beta = +0.006$, and $r^{(0)} = 0.4a^{(0)}$.

**Table I**

|  | General relativity | Effective field theory of distorted photonic-crystals |
|---|---|---|
| Source of gravity | Matter | Lattice distortion |
| Field Action | Einstein – Hilbert Action | None |
| Equation of Motion | Einstein Equation | Distortion tensor |
| Gravitation Field | $g_{\mu\nu}$ (Solution) | $g_{\mu\nu}$ |
| Fermat's Action | $\int ds$ | $\int ds$ |
| Master equation of light | geodesics equation | geodesics equation |

**Fig. 1**

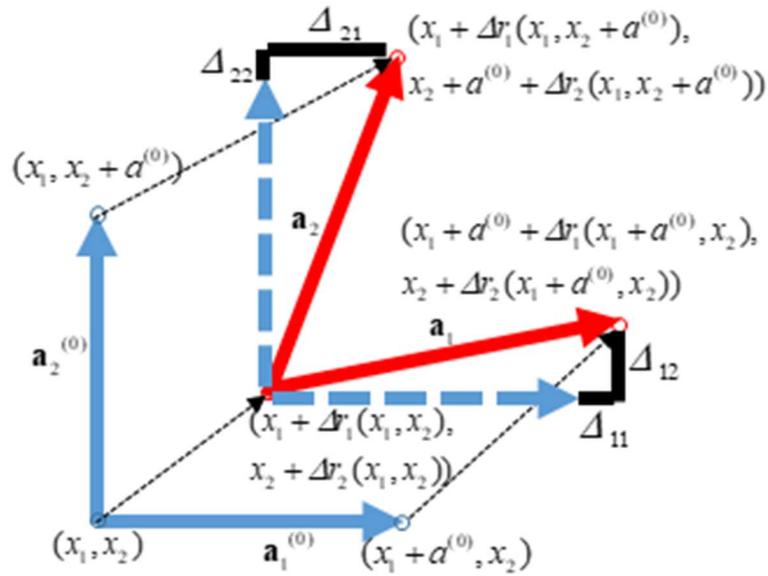

**Fig. 2**

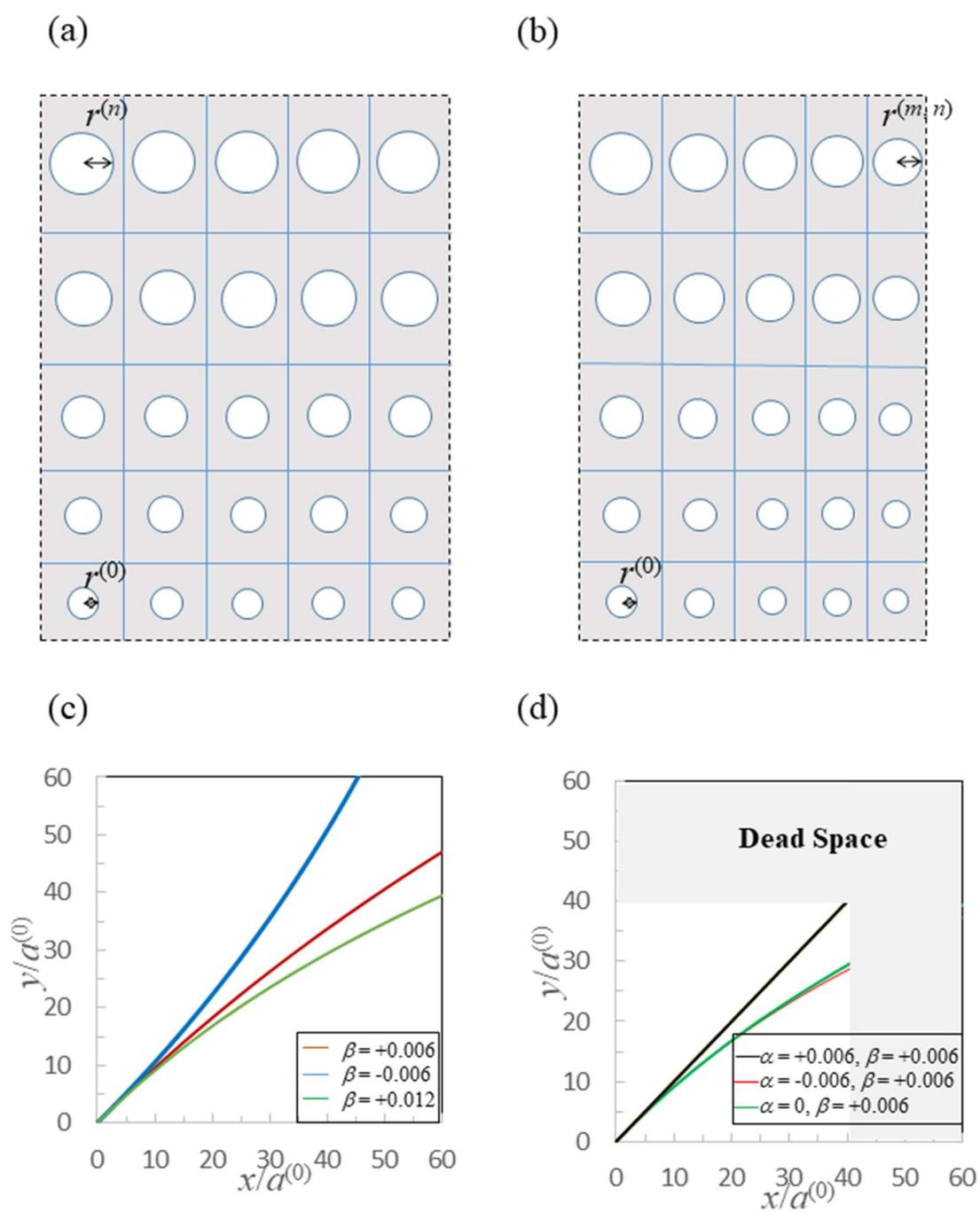

**Fig. 3**

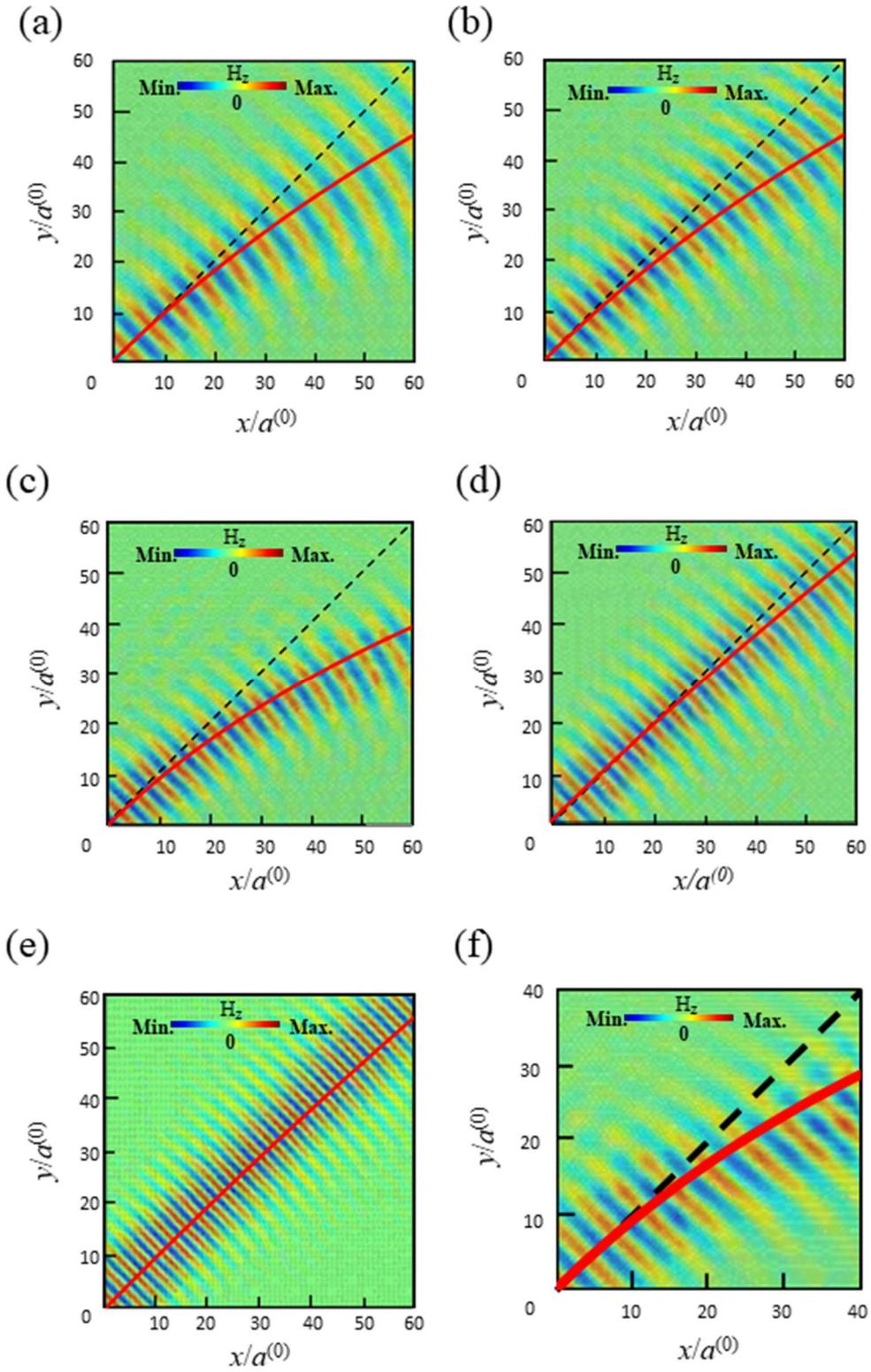